# Seismic signal sparse time-frequency analysis by Lp-quasinorm constraint

Yingpin Chen, Zhenming Peng*, Member, IEEE, Ali Gholami, Jingwen Yan and Shu Li

*Abstract*—Time-frequency analysis has been applied successfully in many fields. However, the traditional methods, like short time Fourier transform and Cohen distribution, suffer from the low resolution or the interference of the cross terms. To solve these issues, we put forward a new sparse time-frequency analysis model by using the Lp-quasinorm constraint, which is capable of fitting the sparsity prior knowledge in the frequency domain. In the proposed model, we regard the short time truncated data as the observation of sparse representation and design a dictionary matrix, which builds up the relationship between the short time measurement and the sparse spectrum. Based on the relationship and the Lp-quasinorm feasible domain, the proposed model is established. The alternating direction method of multipliers (ADMM) is adopted to solve the proposed model. Experiments are then conducted on several theoretical signals and applied to the seismic signal spectrum decomposition, indicating that the proposed method is able to obtain a higher time-frequency distribution than state-of-the-art time-frequency methods. Thus, the proposed method is of great importance to reservoir exploration.

*Index Terms*—ADMM, sparse representation, Lp-quasinorm constraint, seismic signal decomposition, sparse time-frequency analysis.

## I. INTRODUCTION

IN seismic exploration, when the seismic signals go through the field containing the gas or oil, the amplitude of the high-frequency components would decrease significantly. Therefore, the time-frequency analysis is one of the key technologies to predict the position of the reservoir. At present, the time-frequency analysis technologies used in seismic exploration can be mainly divided into two categories. One is the linear time-frequency analysis technology, including short-time Fourier transform (STFT) [1], wavelet transform [2], S transform [3] and Chirplet transform [4], *etc*. The linear time-frequency analysis technology is calculated efficiently but suffered from low resolution because of the constraint of the Heisenberg uncertainty principle [1]. The other is the bilinear time-frequency analysis technology, which is summarized as a uniform form called Cohen distribution [5]. Cohen distribution suffered from the interference of cross terms [6]. Considering the drawbacks of the linear and bilinear time-frequency technologies, many efforts are made to improve the traditional time-frequency technologies [7-14]. These efforts mainly focus on obtaining a higher resolution time-frequency distribution (TFD).

Recently, sparse representation (SR) [15-17] has become a new branch of technology in signal processing. Time-frequency analysis technologies based on SR have appeared. For example, Liu *et al.* [16] proposed a short time sparse representation algorithm based on the smooth L0-norm algorithm (STSR-SL0). Stanković *et al.* [18] regarded the polynomial Fourier transform as a sparse domain and obtained a sparse time-frequency distribution by the sparse representation in a sparse domain based on L1-norm. Flandrin and Borgnat [19] introduced the compressed sensing (CS) [20] into bilinear time-frequency analysis and proposed the sparse Cohen distribution (SCD) based on the under-sampling in the ambiguity domain and the L1-norm constraint. Because of the excellent performance of SCD, Whitelonis and Ling [21] applied the SCD to radar signature analysis. However, in the framework of SCD, a square kernel is adopted to obtain the under-sampling data in the ambiguity domain. As a result, the cross term interference may not be eliminated completely [19, 22]. Jokanovic *et al.* [23] improved the SCD by using the adaptive optimal kernel method [11]. Gholami [24] proposed a sparse time-frequency decomposition (STFD) by using the split Bregman iterations. In the framework of STFD, the Kronecker product operator is used. Therefore, a large scale matrix is inevitable, resulting in the heavy calculation. Hu *et al.* [25] added a L2-norm regularization based on the work of Gholami to adjust the sparsity of the spectrum. Wang *et al.* [26] proposed the sparse S transform by using L1-norm. In summary, the sparse time-frequency analysis methods are mainly based on L1-norm constraint. However, the L1-norm is only the convex relaxation of L0-norm [27]. The $p^{th}$ pow of Lp-norm ( $\left\| \boldsymbol{x}_i \right\|_p^p = \sum_{i=1}^{N} \left| \boldsymbol{x}_i \right|^p$ ,

*The corresponding author of this paper is Zhenming Peng.

Yingpin Chen is a Ph.D. candidate of the School of Information and Communication Engineering, University of Electronic Science and Technology of China, Chengdu 610054 China (e-mail: 110500617@163.com).

Zhenming Peng is now the professor and doctor's supervisor of University of Electronic Science and Technology of China, Chengdu 610054, China (e-mail: zmpeng@uestc.edu.cn). He is also the professor of Center for Information Geoscience, University of Electronic Science and Technology of China, Chengdu, China. He is the corresponding author of this paper.

Ali Gholami is with the Institute of Geophysics, University of Tehran, Tehran 14155/6466, Iran (e-mail: agholami@ut.ac.ir).

Jingwen Yan is now the professor and doctor's supervisor of Shantou University, Shantou 515063, China (e-mail: jwyan@stu.edu.cn).

Shu Li is a Ph.D. candidate of the School of Information and Communication Engineering, University of Electronic Science and Technology of China, Chengdu 610054 China (e-mail: lishuvip@126.com).

$0 \leq p \leq 1$, for simplicity, we name $\|x_i\|_p^p$ as Lp-quasinorm) is another relaxation of L0-norm. In fact, L1-norm constraint is a particular case of Lp-quasinorm. Gribonval and Nielsen [28] proved the uniqueness of the solution to Lp- quasinorm based sparse representation (LpSR) theoretically. Recently, Woodworth and Chartrand [29] pointed out that the Lp-quasinorm is better able to approximate the original L0-norm than L1-norm and developed an iterative Lp-quasinorm shrinkage (LpS) solving LpSR. Woodworth and Chartrand also proved the convergence of iterative LpS for solving LpSR. Then, the generalized shrinkage operator was applied in sparse representation. For instance, Zhang *et al.* [30] applied the LpS operator to computed tomography image reconstruction which provided a better quality of reconstruction than the one by using L1-norm constraint. Zuo *et al.* [31] adopted the LpS instead of L1-norm based soft-thresholding shrinkage in image blind deconvolution and obtained a satisfactory performance. Since Lp-quasinorm is the approximation of the L0-norm with a degree of freedom, which can obtain a more precise solution than the L1-norm, we put forward a Lp-quasinorm based local time inversion model in the frequency domain. The advantages of Lp-quasinorm regularization are listed as follows. 1) The LpS operator may lead to a precise and convergent solution. 2) Lp-quasinorm is more flexible than L1-norm. This may fit the degree of sparsity with respect to the processed signal. 3) The Lp-quasinorm feasible domain makes the solution robust to noise.

The proposed model absorbs the truncating thought of STFT, that is, cutting off the signal by a sliding window. In this way, the size of the involved matrix would be very small, which may decrease the amount of calculation. To reduce the interference of the false frequency introduced by the surrounding data in the proposed model, we adopt Gaussian window as the sliding window. Regarding the short time measurement as a part of signal reconstructed by sparse spectrum, the relationship between the short time measurement and the sparse spectrum is then established. Based on this relationship, the Lp-quasinorm regularization is adopted to the proposed model, fitting the prior sparsity information. To solve the proposed model, the alternating direction method of multipliers (ADMM) [32] is adopted. In order to evaluate the performance of the proposed method, experiments based on several theoretical signal data and real seismic signals are conducted by comparing with some state-of-the-art time-frequency analysis methods. The indicators we adopt are the peak signal to noise ratio (PSNR), Renyi entropy, concentration measurement (CM) , relative error (RE) and cost time [33-35]. It can be found from the experiments that the proposed method has the capacity of obtaining an accurate time-frequency representation. The main contributions are listed as follows. 1) We put forward a new time-frequency inversion model, which is a general framework for short time spectrum inversion problem. Therefore, the other sparse representation methods can be easily adopted in the proposed model. 2) The LpS is adopted to the proposed model, which can obtain a precise TFD. 3) Considering the form of the proposed model, we solve the model by ADMM, which provides a precise solution of the model.

The rest of the paper is organized as follows. In Section II, we provide some preliminary knowledge with regard to SR. In Section III, we introduce SR into STFT and put forward a Lp-quasinorm regularization based time-frequency inversion model and discuss how to solve the model by ADMM in detail. In Section IV, based on three theoretical signal data, the experiments are conducted to compare with the traditional methods and some state-of-the-art sparse time-frequency methods. In Section V, the proposed method is applied to the field of reservoir exploration. Finally, we summarize this paper in Section VI.

## II. PRELIMINARY

Sparse representation is a very active branch of signal processing. A diagrammatic sketch of sparse representation is shown in Fig. 1. $y \in \mathbb{C}^{M \times 1}$ is the observation of SR. $\Theta \in \mathbb{C}^{M \times N}(M < N)$ is the dictionary of SR. $x \in \mathbb{C}^{N \times 1}$ is the sparse representation coefficient in which the number of non-zero entries is small. It is shown in Fig. 1 that there are only two non-zero entries. Clearly, to represent $y$, only the second and fifth columns of the dictionary, which are boxed in red, are selected.

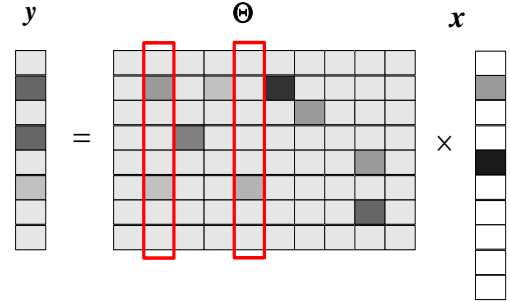

Fig. 1. The diagrammatic sketch of the sparse representation.

The SR framework can be modeled as the $P_0$ issue shown in (1)

$$P_0 : \min \|x\|_0, \quad s.t. \quad y = \Theta x, \tag{1}$$

where L0-norm $\|x\|_0$ is defined as the number of the non-zero entries of $x$ .

Since the issue described in (1) is a non-deterministic polynomial hard (NP-Hard) problem, people often relax the $P_0$ issue as the $P_1$ issue shown in (2).

$$P_1 : \min \|x\|_1, \quad s.t. \quad y = \Theta x, \tag{2}$$

where $\|x\|_1 = \sum_{i=1}^{N} |x_i|$ is the L1-norm of $x$ .

## III. PROPOSED METHOD

### A. The sparse time-frequency inversion model

Signals in the real world are non-stationary. However, the truncated signals are assumed to be stationary. Therefore, the truncated signal weighted by a Gaussian window can be regarded as approximatively sparse in the frequency domain. Our motivation is to establish a relationship between SR and STFT and obtain a high-resolution time-frequency distribution. In addition, as we establish the framework based on the short time measurement, there is no cross-term interference in the proposed time-frequency inversion model.

To introduce SR into STFT, we truncate the original signal $s \in \mathbb{C}^{N \times 1}$ as the sub-signals $s_i \in \mathbb{C}^{M \times 1}$ $(i = 1, 2, \cdots, N)$ where $M$ is an odd number denoting the length of the window. In order to ensure that the length of every sub signal is $M$, we need to pad the original signal on the right and left boundaries with $(M-1)/2$ zeros. After padding, the $(M-1)/2+i$ -th entry of the padding signal is regarded as the center of $s_i$. Then, we truncate the padding signal by extracting the center of $s_i$ and the $(M-1)/2$ entries on both sides of the center from the padding signal. After padding and truncating, $s_i$ is obtained. In order to reduce the influence of the data far away from the center of $s_i$ and enhance the effect of data near the center, a Gaussian window $g \in \mathbb{R}^{M \times 1}$ is needed to reweight $s_i$, that is

$$y_i = g \circ s_i, \qquad (3)$$

where the symbol $\circ$ denotes the componentwise multiplication.

Suppose $x_i \in \mathbb{C}^{N \times 1}$ is the sparse spectrum with respect to $y_i$, thus, the corresponding signal in the time domain is $F^{-1} x_i \in \mathbb{C}^{N \times 1}$ where $F^{-1} \in \mathbb{C}^{N \times N}$ denotes the inverse Fourier transform matrix. However, the size of observation vector $y_i$ is $y_i \in \mathbb{C}^{M \times 1}$, thus, a matrix is required to establish the relationship between $y_i$ and $F^{-1} x_i$. $F^{-1} x_i \in \mathbb{C}^{N \times 1}$ is the reconstructed signal in time domain, which is assumed to an approximation of the observed measurement $y_i$. Regarding $y_i$ as the first $M$ entries of $F^{-1} x_i$, consequently, $y_i$ and $x_i$ can be modeled as

$$y_i = S F^{-1} x_i, \qquad (4)$$

where $S \in \mathbb{R}^{M \times N}$ is the selecting matrix, which is defined as

$$S = \begin{bmatrix} I & | & \mathbf{O} \end{bmatrix}, \qquad (5)$$

where $I \in \mathbb{R}^{M \times M}$ is the identity matrix and $\mathbf{O} \in \mathbb{R}^{M \times (N-M)}$ is the zero matrix whose entries are all zeros.

The $F$ in (4) is the Fourier transform matrix defined as

$$F = \begin{bmatrix} 1 & 1 & 1 & \cdots & 1 \\ 1 & W_N^1 & W_N^2 & \cdots & W_N^{N-1} \\ 1 & W_N^2 & W_N^4 & \cdots & W_N^{2(N-1)} \\ \vdots & \vdots & \vdots & \ddots & \vdots \\ 1 & W_N^{N-1} & W_N^{2(N-1)} & \cdots & W_N^{(N-1) \times (N-1)} \end{bmatrix}, \qquad (6)$$

where $W_N = \exp(-j \dfrac{2\pi}{N})$.

In a traditional STFT framework, after truncation and reweighting, the length of $y_i$ is $M$. However, the size of the spectrum is $N$. Therefore, a zero-padding operation is carried out on $y_i$. Therefore, extra frequency components are brought into STFT. To avoid extra frequency components, a reverse idea is created. In fact, the truncation observation in STFT can be regarded as the observation in SR. Comparing the definition of $\Theta$ with (4), we can find that the dictionary $\Theta = S F^{-1}$. $S$ plays a role of selecting the first $M$ entries of $F^{-1} x_i$.

To illustrate the differences between L1-norm and Lp-quasinorm, we show the contour line of Lp-quasinorm in Fig. 2. As is shown in Fig. 2, the smaller the parameter $p$ is, the sparser the feasible domain of Lp-quasinorm is. In Fig. 3, we demonstrate the advantages of Lp-quasinorm. Suppose the signal is disturbed by Gaussian noise (the standard deviation is $\sigma$) as seen in Fig. 3(a). Obviously, the dotted line in Fig. 3(c) would intersect the contour line of Lp-quasinorm near the axes, inducing a sparser solution (which is also robust to noise) than the one of L1-norm and L2-norm. As is mentioned, Lp-quasinorm may be a better choice. Therefore, we establish the model as (7).

$$P_p: \min \|x_i\|_p^p, \quad s.t. \quad y_i = \Theta x_i \big|_{\Theta = S F^{-1}}, \qquad (7)$$

where $\|x_i\|_p$ is the Lp-norm defined as $\|x\|_p = (\sum_{i=1}^{N} |x_i|^p)^{1/p}$.

In the proposed model, the local sparse frequency is the inverse objective, while we only focus on the similarity between the local measurement and the observation interval of the objective in the time domain. Thus, extra frequency components would not be brought in the reconstructed spectrum.

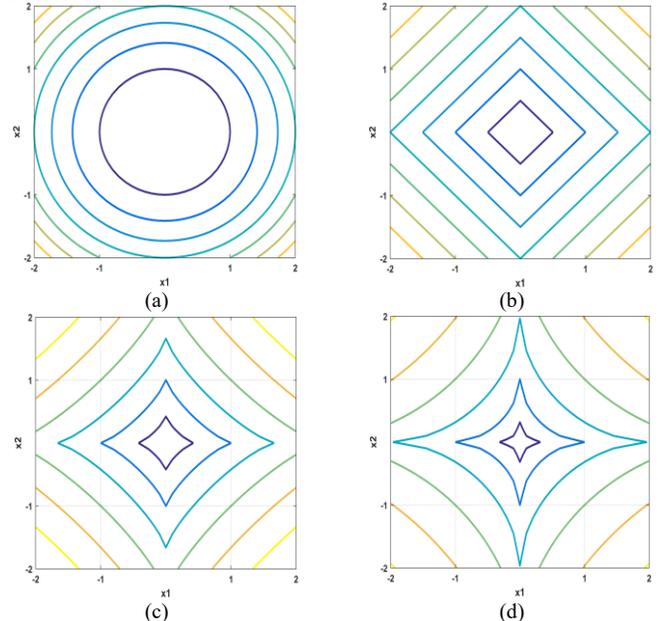

Fig. 2. The contour line of Lp-quasinorm. (a) p=2; (b) p=1; (c) p=0.8; (d) p=0.6.

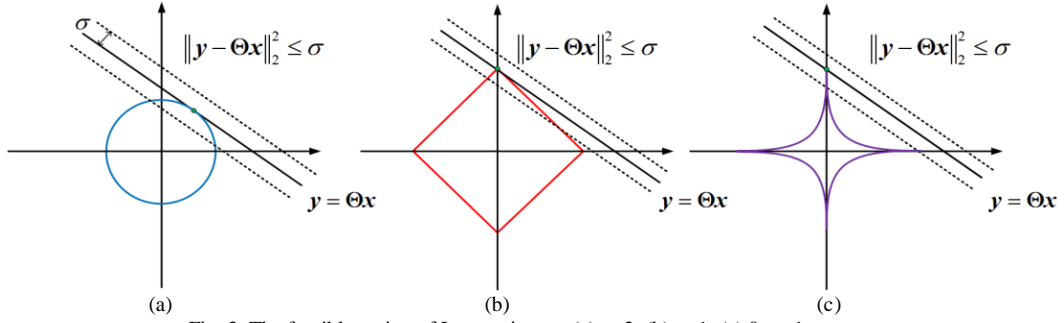

Fig. 3. The feasible region of Lp-quasinorm. (a) p=2; (b) p=1; (c) 0<p<1.

### B. The solver based on ADMM and LpS operator

The issue shown in (7) is a constrained problem, which is equivalent to the unconstrained form shown in (8),

$$\min_{x_i} \{ \mu \|x_i\|_p^p + \frac{1}{2}\|y_i - \Theta x_i\|_2^2 \}, \qquad (8)$$

where $\mu$ is the coefficient, which balances the Lp-quasinorm regularization and the fidelity term $\frac{1}{2}\|y_i - \Theta x_i\|_2^2$.

To solve (8) by ADMM, the splitting variable $z = x_i$ and the Lagrange coefficient $\lambda$ are required. Then we have the augmented Lagrangian function as

$$
\begin{aligned}
J = \max_{\lambda} \min_{x_i, z} \{ &\mu \|z\|_p^p + \frac{1}{2}\|y_i - \Theta x_i\|_2^2 \\
&- \langle \lambda, z - x_i \rangle + \frac{\beta}{2}\|z - x_i\|_2^2 \},
\end{aligned}
\qquad (9)
$$

where $\beta > 0$ is the penalty coefficient.

For the convenience of completing the square in subsequent calculations, we set $\lambda = \beta \tilde{z}$, then (9) becomes (10).

$$
\begin{aligned}
J = \max_{\tilde{z}} \min_{x_i, z} \{ &\mu \|z\|_p^p + \frac{1}{2}\|y_i - \Theta x_i\|_2^2 \\
&- \beta \langle \tilde{z}, z - x_i \rangle + \frac{\beta}{2}\|z - x_i\|_2^2 \}.
\end{aligned}
\qquad (10)
$$

With respect to the $x_i$ subproblem, the sub-objective function is

$$J_{x_i} = \min_{x_i} \{ \frac{1}{2}\|y_i - \Theta x_i\|_2^2 + \frac{\beta}{2}\|z^{(k)} - x_i - \tilde{z}^{(k)}\|_2^2 \}. \quad (11)$$

Setting the first-order derivative of $J_{x_i}$ with regard to $x_i$ zero, that is, $\frac{\partial J_{x_i}}{\partial x_i} = 0$, then we have

$$x_i^{(k+1)} = (\Theta^H \Theta + \beta J)^{-1}(\Theta^H y_i + \beta(z^{(k)} - \tilde{z}^{(k)})), \quad (12)$$

where $J \in \mathbb{R}^{N \times N}$ is the identity matrix.

It should be pointed out that $\Theta^H \Theta + \beta J \in \mathbb{C}^{N \times N}$ is a large scale matrix and the computation complexity of calculating $(\Theta^H \Theta + \beta J)^{-1}$ directly is $O(N^3)$. To decrease the computation complexity of $(\Theta^H \Theta + \beta J)^{-1}$, the following theorem is required.

**Theorem 3.1** : with respect to any matrix $A \in \mathbb{C}^{N \times K}$, $B \in \mathbb{C}^{N \times M}$, $C \in \mathbb{C}^{M \times K}$, the matrix $(A + BC)^{-1}$ satisfies the following equation [36],

$$(A + BC)^{-1} = A^{-1} - A^{-1}B(I + CA^{-1}B)^{-1}CA^{-1}, \quad (13)$$

where $I \in \mathbb{R}^{M \times M}$ is the identity matrix.

Let $A = \beta J$, $B = \Theta^H$, $C = \Theta$. Obviously, $(\Theta^H \Theta + \beta I_1)^{-1}$ is equal to the expression shown in (14).

$$(\Theta^H \Theta + \beta J)^{-1} = \beta^{-1}J - \beta^{-2}\Theta^H (I + \Theta\beta^{-1}\Theta^H)^{-1}\Theta. \quad (14)$$

Since $M$ is usually much smaller than $N$, as a result, the computation complexity of $(I + \beta^{-1}\Theta\Theta^H)^{-1} \in \mathbb{C}^{M \times M}$ becomes $O(M^3)$. Therefore, the computation complexity of $(\Theta^H \Theta + \beta J)^{-1}$ becomes $O(N^2 M)$. In this way, we reduce the computation complexity of the whole algorithm remarkably.

With respect to the $z$ sub-problem, fixing $\tilde{z}$ and $x_i$, the sub-objective function becomes

$$J_z = \min_z \{ \mu \|z\|_p^p - \beta \langle \tilde{z}^{(k+1)}, z - x_i^{(k+1)} \rangle + \frac{\beta}{2}\|z - x_i^{(k+1)}\|_2^2 \}. \quad (15)$$

By using the method of completing square, then we have

$$
\begin{aligned}
z^{(k+1)} &= \arg\min_z \{ \mu \|z\|_p^p - \beta \langle \tilde{z}^{(k)}, z - x_i^{(k+1)} \rangle + \frac{\beta}{2}\|z - x_i^{(k+1)}\|_2^2 \} \\
&= \arg\min_z \{ \mu \|z\|_p^p - \beta \langle \tilde{z}^{(k)}, z - x_i^{(k+1)} \rangle + \frac{\beta}{2}\|z - x_i^{(k+1)}\|_2^2 \\
&\quad + \beta(\tilde{z}^{(k)})^2 - \beta(\tilde{z}^{(k)})^2 \} \\
&= \arg\min_z \{ \mu \|z\|_p^p + \frac{\beta}{2}\|z - x_i^{(k+1)} - \tilde{z}^{(k)}\|_2^2 - \beta(\tilde{z}^{(k)})^2 \} \\
&= \arg\min_z \{ \mu \|z\|_p^p + \frac{\beta}{2}\|z - x_i^{(k+1)} - \tilde{z}^{(k)}\|_2^2 \}.
\end{aligned}
\qquad (16)
$$

The solution of (16) is

$$z^{(k+1)} = shrink_p(x_i^{(k+1)} + \tilde{z}^{(k)}, \frac{\mu}{\beta}), \quad (17)$$

With regard to real number, Lp shrinkage operator is [29] [37]

$$shrink_p(\xi, \tau) = \max \{ |\xi| - \tau^{2-p}|\xi|^{p-1}, 0 \} \frac{\xi}{|\xi|}. \quad (18)$$

where $\tau$ is the threshold. When $p=1$, the LpS degrades as the soft-thresholding shrinkage.

Considering that the data in the time-frequency distribution is complex number, we rewrite the Lp shrinkage operator as

$$shrink_p(\boldsymbol{\xi},\tau)=\max\left\{|\boldsymbol{\xi}|-\tau^{2-p}|\boldsymbol{\xi}|^{p-1},0\right\}\exp(j\varphi(\boldsymbol{\xi})), \quad (19)$$

where $\varphi(\boldsymbol{\xi})$ is the phase of $\boldsymbol{\xi}$.

With regard to $\tilde{z}$, the sub-objective function is

$$J_{\tilde{z}}=\max_{\tilde{z}}\beta\left\langle\tilde{z},\boldsymbol{x}_i^{(k+1)}-\boldsymbol{z}^{(k+1)}\right\rangle. \quad (20)$$

By using the gradient ascent method, the updating rule of $\tilde{z}$ is

$$\tilde{z}^{(k+1)}=\tilde{z}^{(k)}+\gamma\beta(\boldsymbol{x}_i^{(k+1)}-\boldsymbol{z}^{(k+1)}), \quad (21)$$

where $\gamma$ is the learning rate.

Considering the form of Fourier transform matrix shown in (6), a frequency shifting operator is required, that is

$$\boldsymbol{x}_i^{(k+1)}=fftshift(\boldsymbol{x}_i^{(k+1)}) \quad (i=1,2,\cdots,N). \quad (22)$$

where $fftshift$ denotes the centering operator, which exchanges the position of the first half and the second half parts of $\boldsymbol{x}_i^{(k+1)}$.

In summary, the proposed method, which is called sparse time-frequency analysis via LpS (STFA-LpS), is summarized as Algorithm 1.

---

**Algorithm 1 STFA-LpS**

---

**Input:** $s$, $p$, $\boldsymbol{\Theta}$.

**Output:** $\boldsymbol{x}_i$ $(i=1,2,\cdots,N)$.

**Initialize:**

$\boldsymbol{x}_i^{(1)}=\boldsymbol{0}$, $\boldsymbol{z}^{(1)}=\boldsymbol{0}$, $\tilde{z}^{(1)}=\boldsymbol{0}$, $\beta$, $\mu$, $M$, $N$, $\gamma$, $Max$.

1: Truncate the sub-signal $\boldsymbol{s}_i$ $(i=1,2,\cdots,N)$ from the processed signal and weight the sub-signal as shown in (3);

2: **Repeat**

3: **For** $k=1:Max$ **do**

4: $\boldsymbol{x}_i^{(k+1)}=(\boldsymbol{\Theta}^H\boldsymbol{\Theta}+\beta\boldsymbol{J})^{-1}(\boldsymbol{\Theta}^H\boldsymbol{y}_i+\beta(\boldsymbol{z}^{(k)}-\tilde{z}^{(k)}))$;

5: $\boldsymbol{z}^{(k+1)}=shrink_p(\boldsymbol{x}_i^{(k+1)}+\tilde{z}^{(k)},\mu/\beta)$;

6: $\tilde{z}^{(k+1)}=\tilde{z}^{(k)}+\gamma\beta(\boldsymbol{x}_i^{(k+1)}-\boldsymbol{z}^{(k+1)})$;

7: **End for**

8: $i=i+1$;

9: **Until** every weighted sub-signal $\boldsymbol{y}_i$ is processed;

10: $\boldsymbol{x}_i^{(k+1)}=fftshift(\boldsymbol{x}_i^{(k+1)})$ $(i=1,2,\cdots,N)$;

11: **Return** $\boldsymbol{x}_i^{(k+1)}$ $(i=1,2,\cdots,N)$ **as** $\boldsymbol{x}_i$ $(i=1,2,\cdots,N)$.

---

where the symbol $Max$ denotes the maximal iteration number of the iterations. In our experiments, the parameters are set as $p=0.1$, $\beta=1$, $\mu=0.5$, $\gamma=1$, $M=11$, $Max=25$. It is worth pointing out that all the sub-signals can be processed in a parallel fashion.

## IV. NUMERICAL EXPERIMENTS

### A. The running environment and evaluating indicators

The experiments are all implemented in MATLAB 2016 and executed on a laptop equipped with an Intel Core i5-3210M 2.5GHz CPU and 6G of RAM. In this section, we evaluate the proposed method by general measurements, which are the concentration measurement (CM), Renyi entropy, peak signal to noise ratio (PSNR), relative error (RE) and cost time.

$$CM=\frac{\int_{-\infty}^{\infty}\int_{-\infty}^{\infty}|TF(t,f)|^4 dt df}{\left(\int_{-\infty}^{\infty}\int_{-\infty}^{\infty}|TF(t,f)|^2 dt df\right)^2}, \quad (23)$$

where the symbol $TF(t,f)$ represents the time-frequency distribution. The larger the CM is, the better the quality of TFD is.

The Renyi entropy shown in (24) is adopted to measure the aggregation of TFD.

$$R_\alpha(TF(t,f))=\frac{1}{1-\alpha}\log_2\int_{-\infty}^{\infty}\int_{-\infty}^{\infty}TF^\alpha(t,f)dt df, \quad (24)$$

where $\alpha=3$. The smaller the $R_\alpha$ is, the better the quality of TFD is.

In this paper, we carry out experiments on three ideal model, whose ideal time-frequency distribution is known. Therefore, the PSNR can be used to evaluate the methods. All the TFDs are normalized to the range [0,1] and then compared with the ideal TFD by PSNR. The PSNR is defined as

$$PSNR(\boldsymbol{X},\boldsymbol{Y})=10\lg\frac{MAX(\boldsymbol{Y})^2}{\frac{1}{N^2}\sum_{i=1}^{N}\sum_{j=1}^{N}(\boldsymbol{X}_{ij}-\boldsymbol{Y}_{ij})^2}, \quad (25)$$

where $\boldsymbol{Y}$ represents the ideal time-frequency image and the $\boldsymbol{X}$ is the time-frequency distribution obtained by certain time-frequency analysis algorithm. The larger the PSNR is, the greater the similarity of $\boldsymbol{X}$ and $\boldsymbol{Y}$ is.

The relative error is defined as

$$RE(\boldsymbol{X},\boldsymbol{Y})=\frac{\|\boldsymbol{X}-\boldsymbol{Y}\|_2}{\|\boldsymbol{X}\|_2}. \quad (26)$$

As is defined in (26), the smaller the relative error is, the better the quality of time-frequency distribution is.

### B. Signal 1: mono-component signal

Linear frequency modulated (LFM) signal is often used to measure the quality of time-frequency analysis. The closed form of the tested LFM signal is defined as

$$s(t)=\exp(j\pi t^2), \quad (27)$$

where time range is from -8 s to 8 s and the sampling rate is 16 Hz.

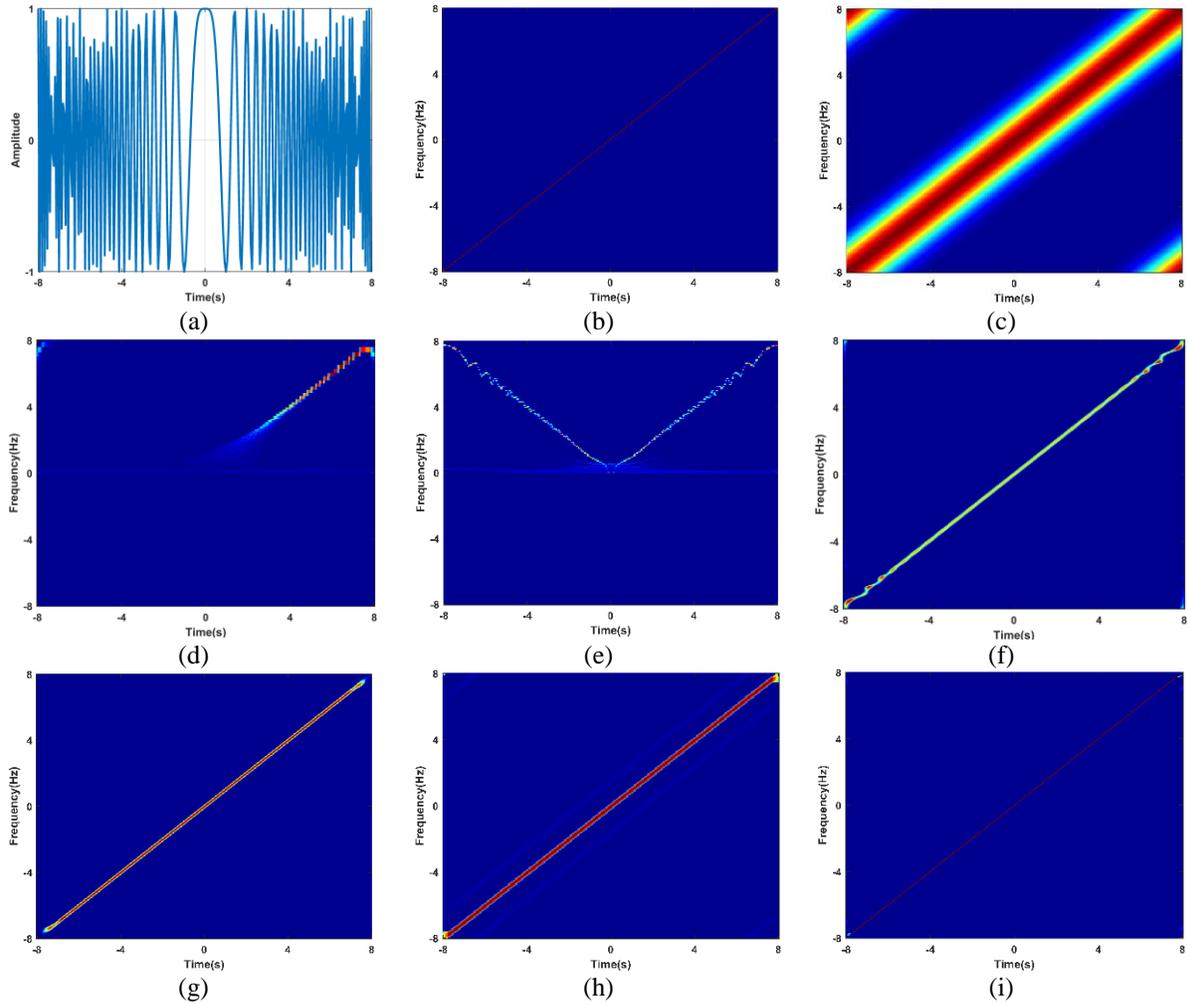

Fig. 4. The real part of the processed signal and the TFDs. (a) The real part of the processed signal; (b) the ideal time-frequency distribution; (c) the result of STFT; (d) the result of SSWT; (e) the result of HHT; (f) the result of STFD; (g) the result of SCD; (h) the result of STSR-SL0; (i) the result of the proposed method.

The instantaneous phase of the LFM signal is $\varphi(t) = \pi t^2$, thus, the instantaneous frequency is

$$f(t) = \frac{1}{2\pi} \frac{d\varphi(t)}{dt} = t. \tag{28}$$

Therefore, the ideal TFD should be a straight line. Table I provides the evaluating indicators of TFDs shown in Fig. 4, where the TFDs are obtained by STFT [1], synchrosqueezed wavelet transform (SSWT) [38], Hilbert-Huang Transform (HHT) [13], STFD [24], SCD [20], STSR-SL0 [16] and the proposed method with regard to signal 1. It is obvious that the TFD obtained by the proposed method has the best performance in PSNR, Renyi entropy, and CM.

As is shown in Fig. 4 that the result of the proposed method is closest to the ideal TFD shown in Fig. 4(b). The other TFDs have some disadvantages shown in Fig. 4. For example, the TFD by STFT does not have a satisfactory time-frequency resolution. Since the SSWT is based on the wavelet transform, the SSWT can only represent the positive frequency components shown in Fig. 4(d). In addition, the SSWT losses the low-frequency information comparing with other methods. Fig. 4(e) shows that HHT can only represent the positive frequency component because of the use of Hilbert transform. To demonstrate the effectiveness of the LpS, we figure out the relative error curve in Fig. 5 with respect to the ideal instantaneous spectrum slice at the time $t = 0$ in Fig. 4 (b).

TABLE I
THE EVALUATING INDICATORS OF TFDS IN FIG. 4

| Algorithm | Linear frequency modulation signal | | | |
|---|---|---|---|---|
| | PSNR(dB) | Renyi | CM | Time(s) |
| STFT | 8.76 | 14.05 | 8.0E-5 | **0.04** |
| SSWT | 23.04 | 9.28 | 4.0E-3 | 0.17 |
| HHT | 22.90 | 9.13 | 3.3E-3 | 0.72 |
| STFD | 23.65 | 10.10 | 1.0E-3 | 8.73 |
| SCD | 22.12 | 9.94 | 1.2E-3 | 2.67 |
| STSR-SL0 | 20.55 | 11.09 | 9.0E-4 | 3.36 |
| STFA-LpS | **36.43** | **8.02** | **4.1E-3** | 1.39 |

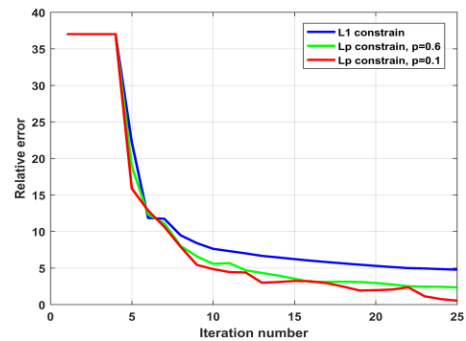

Fig. 5. The relative error curve.

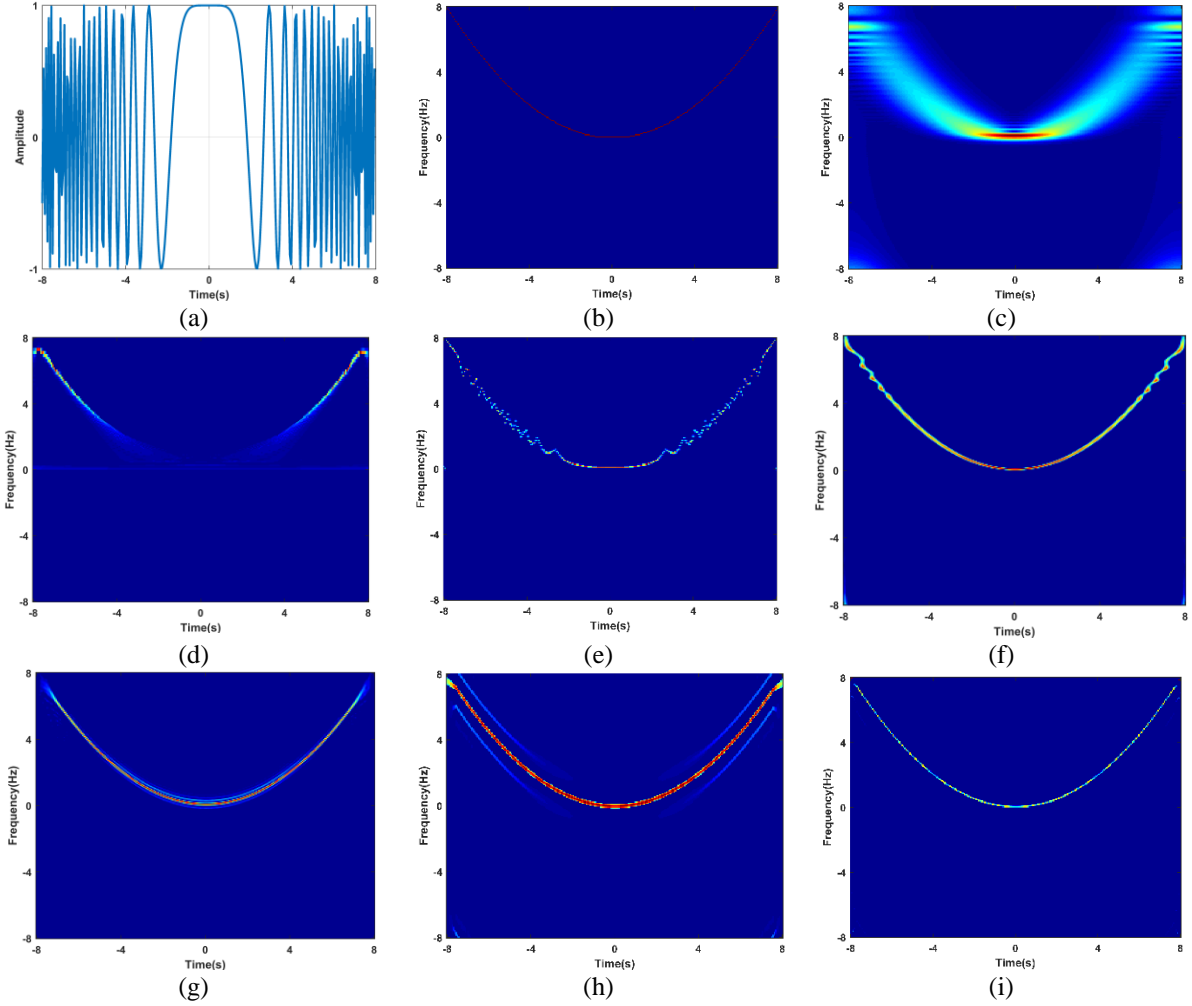

Fig. 6. The real part of the processed signal and the TFDs. (a) The real part of the processed signal; (b) the ideal time-frequency distribution; (c) the result of STFT; (d) the result of SSWT; (e) the result of HHT; (f) the result of STFD; (g) the result of SCD; (h) the result of STSR-SL0; (i) the result of the proposed method.

As is shown, the smaller the $p$ is, the smaller the RE with respect to the ideal TFD is.

### C. Signal 2 : parabola frequency modulated signal

The second signal we adopt is the parabola frequency modulated signal, whose closed form is shown as follow

$$s(t) = \exp(j\frac{\pi t^3}{12}), \tag{29}$$

where the time range is $t \in [-8s, 8s]$ and the sampling rate is 16Hz. The instantaneous phase of the second signal is $\varphi(t) = \frac{\pi t^3}{12}$, consequently, the instantaneous frequency is

$$f(t) = \frac{1}{2\pi}\frac{d\varphi(t)}{dt} = \frac{t^2}{8}. \tag{30}$$

Thus, the shape of the ideal time-frequency distribution with respect to the second signal is a parabola. Therefore, the instantaneous frequency of the signal changes fast. As a result, the short time truncated signal may include many components. In other words, the spectrum of the truncated signal is not sparse.

However, the ideal time-frequency distribution of parabola frequency modulated signal is sparse according to (30). Under this circumstance, the weighted function is required to eliminate the influence of the data near the center of the truncated signal and force the sparse spectrum to be sparse.

With regard to signal 2, we provide the TFDs by STFT, SSWT, HHT, STFD, SCD, STSR-SL0 and the proposed method. The performances of the algorithms are recorded in Table Ⅱ. It can be found that the PSNR and CM obtained by our method are the highest while the Renyi entropy by our method is the smallest. Therefore, the proposed algorithm outperforms the other algorithms.

TABLE Ⅱ
THE EVALUATING INDICATORS OF TFDs IN FIG. 6

| Algorithm | Parabola frequency modulated signal | | | |
|---|---|---|---|---|
| | PSNR(dB) | Renyi | CM | Time(s) |
| STFT | 16.95 | 13.75 | 1.5E-4 | **0.03** |
| SSWT | 23.71 | 9.92 | 3.0E-3 | 0.30 |
| HHT | 22.90 | 9.13 | 3.6E-3 | 0.19 |
| STFD | 22.27 | 10.01 | 1.2E-3 | 8.72 |
| SCD | 19.05 | 10.02 | 1.8E-3 | 2.58 |
| STSR-SL0 | 20.83 | 9.86 | 9.3E-4 | 3.22 |
| STFA-LpS | **23.92** | **7.79** | **1.0E-2** | 1.29 |

Fig. 6 provides the real part of signal 2 and TFDs computed by different methods. In Fig. 6, it is observed that the TFD calculated by the proposed method shown in figure 6(i) is the closest TFD with respect to the ideal TFD shown in Fig. 6(b). On the one hand, by using the Gaussian window, the spectrum becomes very sparse and avoids the influence of the surrounding data like STFT shown in Fig. 6(c). On the other hand, it is not interfered by the cross terms like Fig. 6(g) and the false frequency components like HHT shown in Fig. 6(e). It is observed that the result of SSWT shown in Fig. 6 (d) losses some frequency information among the range $t \in [-4s, 4s]$ while the result of ours depict the time-frequency trajectory precisely.

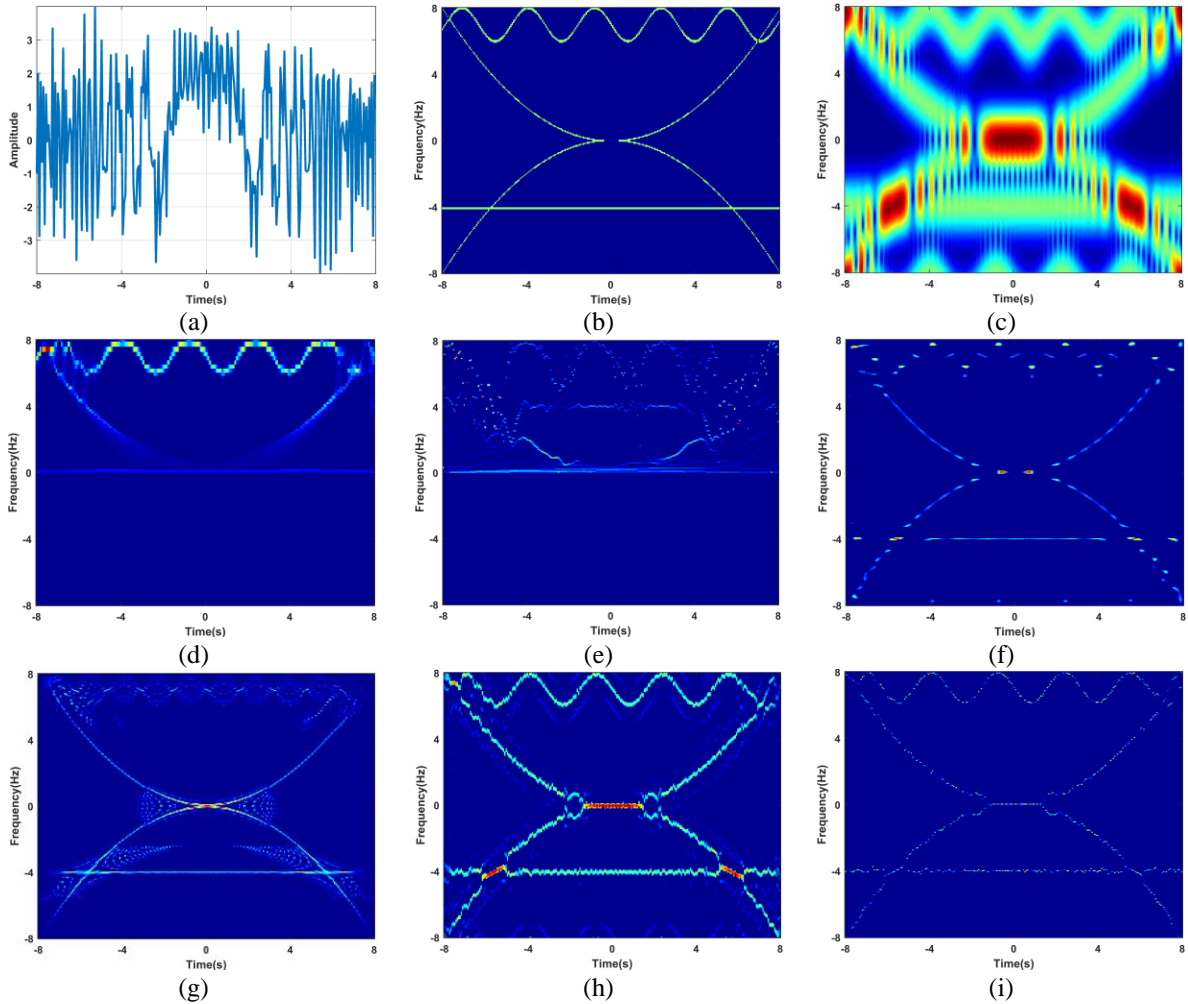

Fig. 7. The real part of the processed signal and the TFDs. (a) The real part of the processed signal; (b) the ideal time-frequency distribution; (c) the result of STFT; (d) the result of SSWT; (e) the result of HHT; (f) the result of STFD; (g) the result of SCD; (h) the result of STSR-SL0; (i) the result of the proposed method.

### D. Signal 3 : multi-component signal

The aforementioned signals are mono-component signals. However, the real signals are always composed of many components. Without loss of generality, we carry out the comparisons on a multi-component signal, which is

$$s(t) = \exp(j\frac{\pi t^3}{12}) + \exp(-j\frac{\pi t^3}{12}) + \exp(-j8\pi t) \quad (31)$$
$$+ \exp(-j\pi(14t + \cos(2t))),$$

where $t \in [-8s, 8s]$ and the sampling rate is 16 Hz.

As is defined in (31), the third signal contains the stationary component $\exp(-j8\pi t)$ , the fast time-varying frequency component $\exp(-j\pi(14t + \cos(2t)))$ and the slow time-varying frequency components $\exp(j\frac{\pi t^3}{12})$ and $\exp(-j\frac{\pi t^3}{12})$ .

With regard to signal 3, we provide the TFDs by STFT, SSWT, HHT, STFD, SCD, STSR-SL0 and the proposed method. The performances of the algorithms are recorded in Table Ⅲ and TFDs are listed in Fig. 7.

TABLE Ⅲ
THE EVALUATING INDICATORS OF TFDS IN FIG. 7

| Algorithm | Multi-component signal | | | |
|---|---|---|---|---|
| | PSNR(dB) | Renyi | CM | Time(s) |
| STFT | 8.45 | 15.31 | 3.67e-5 | **0.05** |
| SSWT | 20.53 | 11.48 | 8.17e-4 | 0.17 |
| HHT | 20.45 | 10.35 | 1.8E-3 | 0.67 |
| STFD | 21.03 | 9.99 | 1.9E-3 | 10.61 |
| SCD | 20.54 | 11.43 | 8.60E-4 | 4.42 |
| STSR-SL0 | 19.96 | 12.30 | 3.7E-4 | 2.75 |
| STFA-LpS | **21.15** | **9.41** | **2.5E-3** | 2.17 |

It is observed that the PSNR, Renyi entropy, and CM of the proposed method with regard to the third signal are still best shown in Table Ⅲ.

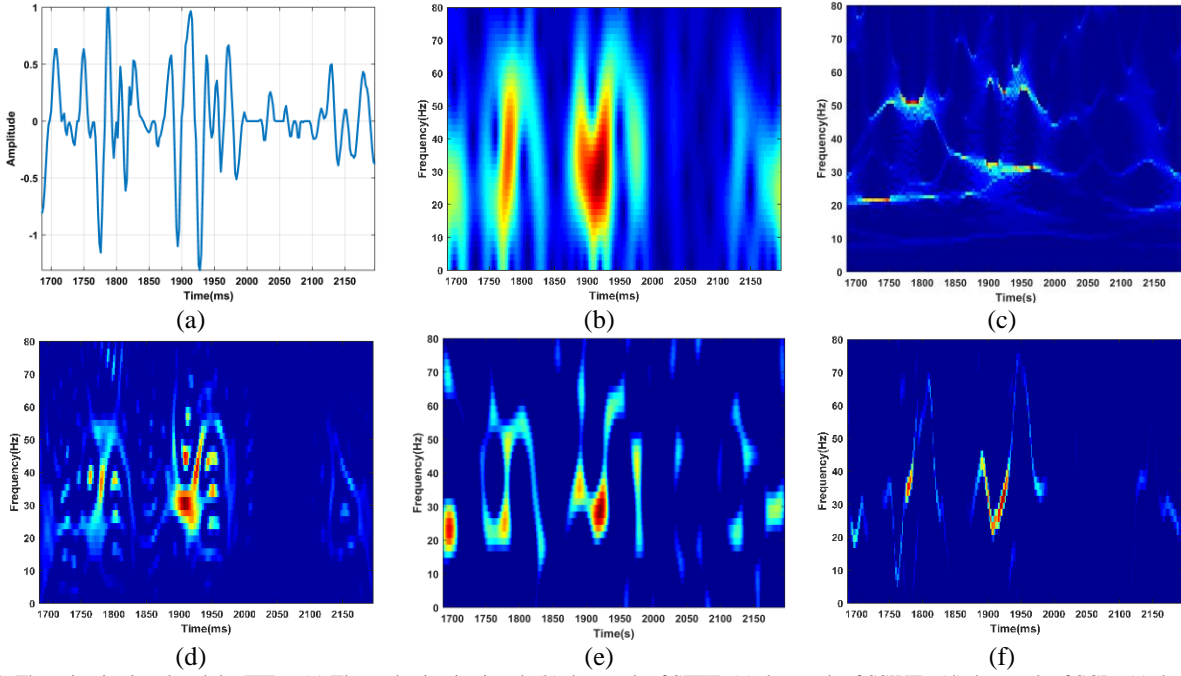

Fig. 8. The seismic signal and the TFDs. (a) The real seismic signal; (b) the result of STFT; (c) the result of SSWT ; (d) the result of SCD; (e) the result of STFD; (f) the result of STFA-LpS.

As is shown in Fig. 7, the result of STFT shown in Fig. 7(c) does not have a high time-frequency resolution. It is observed in Fig. 7(d)-(e) that the result of SSWT and HHT cannot show the negative frequency component. The result of STFD shown in Fig. 7(f) cannot reflect the fast time-varying frequency component. The TFD calculated by SCD is interfered by cross terms shown in Fig. 7(g). In terms of STSR-SL0 shown in Fig. 7(h), because the sliding window of STSR-SL0 is a rectangular window， some false frequency is involved in the result of STFR-SL0. Thus, the TFD by using the proposed method may be the most satisfactory among the TFDs shown in Fig. 7(i).

## V. APPLICATION IN SEISMIC SIGNAL PROCESSING

### A. Experiment on single seismic signal

The seismic signal shown in Fig. 8(a) comes from an oil field of Sichuan Basin, China. To illustrate the performance of our method, we carry out experiments on STFT, SSWT, SCD, STFD, and the proposed method shown in Fig. 8 (b)-(f).

TABLE Ⅳ
THE EVALUATING INDICATORS OF TFDs IN FIG. 8

| Algorithm | The single seismic signal | | |
|---|---|---|---|
| | Renyi | CM | Time(s) |
| STFT | 13.73 | 1.8E-4 | **0.03** |
| SSWT | 11.95 | 1.0E-3 | 0.58 |
| SCD | 9.97 | 2.2E-3 | 2.50 |
| STFD | 10.50 | 1.0E-3 | 3.83 |
| STFA-LpS | **8.12** | **8.3E-3** | 1.66 |

Considering that there is no standard TFD with respect to the seismic signal, we only use Renyi entropy, CM and cost time to measure the performances of TFDs. The performances are recorded in Table Ⅳ. It is worth pointing out that the Renyi entropy by the proposed method is the smallest and CM obtained by our method is the largest, indicating that proposed method can complete with state-of-the-art methods in the quality of time-frequency distribution. In terms of cost time, the sparse time-frequency methods are slower than the traditional methods while the SSWT runs fastest among the sparse time-frequency methods.

As is seen in Fig. 8, the seismic signal is a signal with fast frequency changing. As it is observed in Fig. 8(a), the signal in the range 1850-1950 ms starts with low frequency, then the frequency goes even lower and gets to the lowest frequency point at the time 1910 ms for signal near this area changes slowest. Then in the range 1910-1950 ms, the frequency of signal becomes larger gradually. Thus, the ideal time-frequency trajectory of the seismic signal in the range 1850-1950 ms should start from a low value, then decrease to the lowest position at the time 1910 ms and finally increase to a higher value as the process progressed. To a certain extent, the result of STFT shown in Fig. 8(b) reflects the process. However, the resolution of STFT is unsatisfactory. For the operation of synchrosqueezing, the energy in the TFD of SSWT shown in Fig. 8(c) focuses on several positions, which cannot depict the time-varying frequency process clearly. Observing Fig. 8(d)-(f), the sparse time-frequency methods can reflect the changing tendency clearly. However, since SCD is a kind of Cohen distribution, the TFD of SCD still exists cross terms shown in Fig. 8(d). As is shown in Fig. 8(f), the proposed method outperforms the other sparse TFDs.

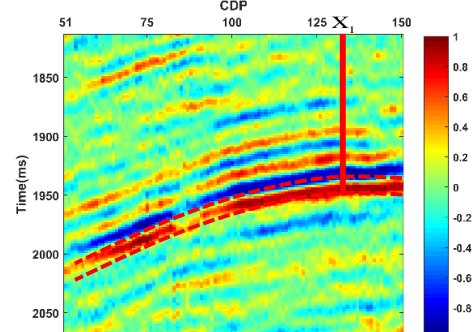

Fig. 9. The seismic signal

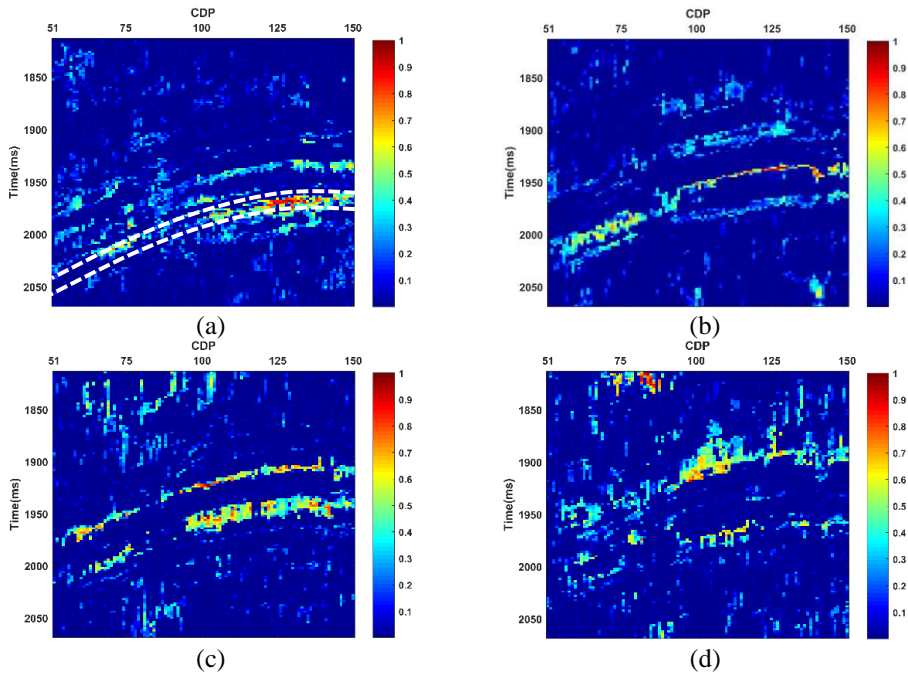

Fig. 10. The single frequency slices obtained by the proposed method. (a) 15 Hz frequency slice; (b) 30 Hz frequency slice; (c) 45 Hz frequency slice; (d) 55 Hz frequency slice.

### B. Application in seismic signal spectral decomposition

In this part, we show the performances of different time-frequency analysis methods based on the two-dimensional seismic signal derived from the Sichuan basin, China. The seismic data is shown in Fig. 9 where the sample rate is 512 Hz. The seismic section from 1814 to 2068 ms is the target stratum for seismic interpretation. In the figure, the time range is 1814–2068 ms, as the ordinate, and the range of common depth point (CDP) is from 51 to 150, as the abscissa. The well $X_1$ near the CDP 131 is a prolific natural gas well, whose daily capacity is 54.18*104 m3/d. According to the exploration information, the location of the well full of gas is 1932-1948 ms. The area full of gas is marked by the red dotted line shown in Fig. 9.

Fig.10 shows the frequency slices calculated by the proposed method. Each slice takes about 30 s. Clearly, the main frequency slice at 30 Hz indicates the location full of gas precisely. It is worth pointing out that the 15 Hz frequency slice shows the low-frequency shadow phenomenon. That is, the amplitude of low-frequency slice below the reservoir is larger than the amplitude on the reservoir. As is shown in Fig. 10 (a), there is a stronger response marked by the white dotted line than the one of the reservoir. The area marked by the white dotted line is below the reservoir. On the contrary, the amplitude of this area becomes smaller than the amplitude of the area full of gas in the 30 Hz slice. The result fits in with the low-frequency shadow phenomenon [39, 40]. In the high-frequency slice shown in Fig. 10(c)-(d), the area containing natural gas has an obvious energy attenuation, which is consistent with high-frequency attenuation in the gas-bearing reservoir.

For comparison, we extract the frequency slices by STSR-SL0. The slices are shown in Fig. 11. Each slice obtained by STSR-SL0 takes about 70 s. It is observed that the slices of each frequency obtained by the proposed method have higher resolution than the ones of STSR-SL0.

In summary, the frequency slices obtained by the proposed method fit the real frequency characteristic of the reservoir without the interference of cross terms. Therefore, the proposed method is capable of being applied to the reservoir exploration because of the high time-frequency resolution.

## VI. CONCLUSION

In this paper, we first reviewed the SR theory, and then, explored the relationship between SR and STFT. Since the short time sampling of STFT can be regarded as the measurement of SR, we build up the local time sparse spectrum construction model. In order to obtain a sparser spectrum, the Lp-quasinorm is used to substitute for the L1 regularization. As the proposed model is established in the framework of STFT, the time-frequency representation by our model is not interfered by the cross terms.

To solve the proposed model, the ADMM framework and the LpS is adopted. As a result, the constrained problem is changed to an unconstrained problem while the variables in the augmented Lagrangian function are all decoupled. Consequently, the proposed model can be solved by several simple sub-problems. For the constraint of the sparsity in the Lp-quasinorm feasible domain, we can easily obtain the time-frequency representation precisely. According to the special.

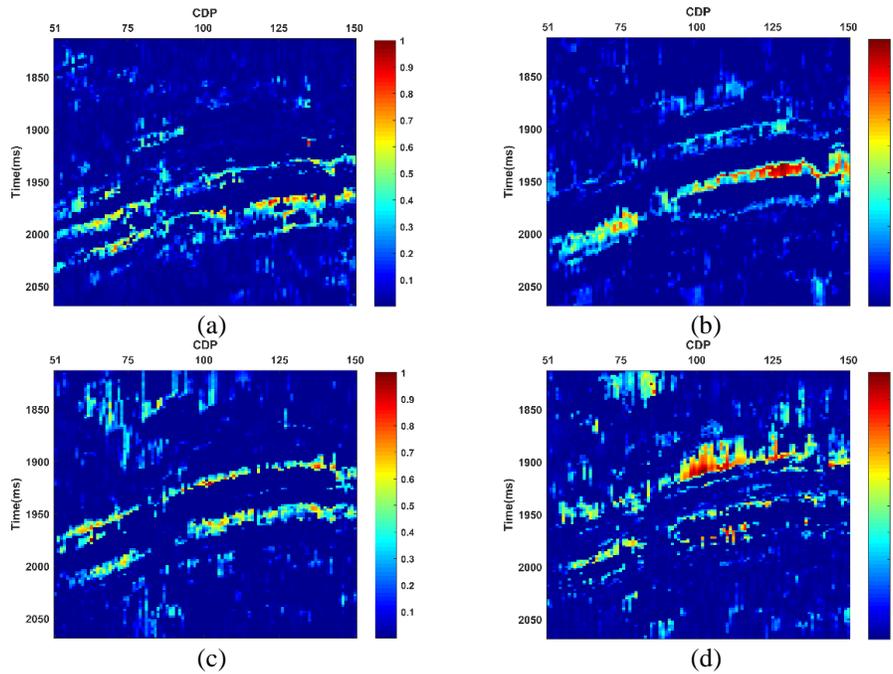

Fig. 11. The single frequency slices obtained by STSR-SL0. (a) 15 Hz frequency slice; (b) 30 Hz frequency slice; (c) 45 Hz frequency slice; (d) 55 Hz frequency slice.

form of the dictionary in the proposed method, we optimized the matrix inversion which reduces the running time strikingly Based on three theoretical signals, we carried out the experiments comparing the proposed method with the other methods and evaluated the methods by PSNR, Renyi entropy, RE, CM, and the cost time. The results showed that the proposed method is capable of calculating high-resolution time-frequency distribution. Finally, the proposed method is applied in the reservoir exploration and obtained high-resolution spectrum decomposition, which is fitting with the low-frequency shadow phenomenon and the high-frequency attenuation.

Although we try our best to optimize the matrix inversion, which can reduce the computational complexity, the proposed method is not as fast as the traditional time-frequency analysis methods. Therefore, we will focus on improving the efficiency of calculating the proposed model in the future.

### ACKNOWLEDGMENT

This work is supported by National Natural Science Foundation of China [grant numbers 61571096, 41274127, 61775030], Natural Science Foundation of Fujian Province [grant number 2015J01270], Education And Scientific Research Foundation of Education Department of Fujian Province for Middle-aged and Young Teachers [grant number JAT170352], and the Open Foundation of Digital Signal and Image Processing Key Laboratory of Guangdong Province [grant number 2017GDDSIPL-01]. We appreciate Dr. Xiaobo Qu of Xiamen University for constructive suggestions that greatly improved the manuscript.

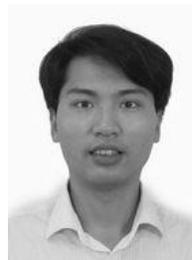

**Yingpin Chen** received his bachelor's degree from Fuzhou University, Fuzhou, Fujian, China, in 2009 while his major is electronic science and technology. He received his master's degree in signal and information processing from University of Electronic and Science Technology of China, Chengdu, Sichuan, China, in 2013. He is now a staff of School of Physics and Information Engineering, Minnan Normal University, Zhangzhou, China. He is now a Ph.D. candidate at University of Electronic and Science Technology of China, Chengdu, Sichuan, China.

His research interests include time-frequency analysis, compressed sensing and convex optimization.

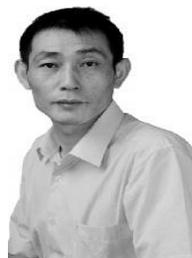

**Zhenming Peng** (M'06) received the Ph.D. degree in geo-detection and information technology from Chengdu University of Technology, Chengdu, China, in 2001. From 2001 to 2003, he was a Postdoctoral Researcher at the Institute of Optics and Electronics, Chinese Academy of Sciences, Chengdu, China. He is currently a professor at University of Electronic Science and Technology of China, Chengdu, China. He is also the professor of Center for Information Geoscience, University of Electronic Science and Technology of China, Chengdu, China.

His research interests include image processing, radar signal processing, and target recognition and tracking.

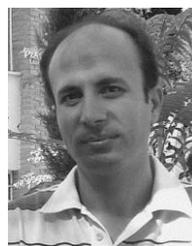

**Ali Gholami** received the Ph.D. degree in geophysics (exploration seismology) from the University of Tehran, Tehran, Iran, in 2010. He is currently an Assistant Professor of geophysics with the Institute of Geophysics, University of Tehran.

His research interests include inverse problems, seismic signal processing, and seismic imaging.

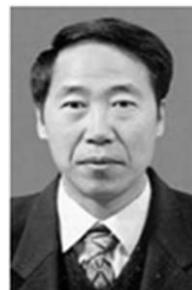

**Jingwen Yan** received his MS's degree from Changchun Geography Institute of Chinese Academy of Sciences and Ph.D. degree from Changchun Geography Institute and Changchun Institute of Optics, Chinese Academy of Sciences, Changchun, Jilin, China. Now, he is a full professor and doctor's supervisor at Shantou University, Shantou, Guangdong, China. His research interests include wavelet analysis and applications, sparse representation of signals, remote sensing, embedded systems, and FPGA/DSP hardware design.

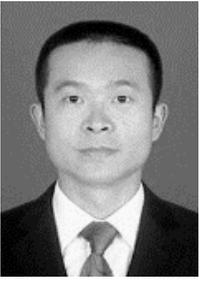 **Shu Li** received his B.S. degree in Communication Engineering from Hebei University, Baoding, Hebei, China, in 2007. He received his master's degree in signal and information processing from Shenzhen University, Shenzhen, Guangdong, China, in 2010 while the major is signal and information processing. He is now a staff of Jishou University, Jishou, Hunan, China. He is now a Ph.D. candidate at University of Electronic and Science Technology of China, Chengdu, Sichuan, China.

His research interests include seismic signal inversion, compressed sensing and convex optimization.